\documentclass[aps,prl,twocolumn,groupedaddress,showpacs]{revtex4}
\usepackage{graphicx,epsf,color,amsmath}
\newif\ifdraft
\newif\ifpreprint
\preprinttrue

\def\fig#1{fig.~{\ref{#1}}}

\def\tab#1{table~{\ref{#1}}}

\def\spa#1.#2{\left\langle#1\,#2\right\rangle}
\def\spb#1.#2{\left[#1\,#2\right]}
\def\tree{{\rm tree}}
\def\Loop{{\rm loop}}

\def\eqn#1{eq.~(\ref{#1})}

\def\NeqFour{{{\cal N}=4}}

\def\NeqEight{{{\cal N}=8}}

\def\be{\begin{equation}}
\def\ee{\end{equation}}
\def\bea{\begin{eqnarray}}
\def\eea{\end{eqnarray}}
\def\ba{\begin{eqnarray}}
\def\ea{\end{eqnarray}}

\def\tree{{\rm tree}}

\newbox\charbox
\newbox\slabox
\def\s#1{{      
        \setbox\charbox=\hbox{$#1$}
        \setbox\slabox=\hbox{$/$}
        \dimen\charbox=\ht\slabox
        \advance\dimen\charbox by -\dp\slabox
        \advance\dimen\charbox by -\ht\charbox
        \advance\dimen\charbox by \dp\charbox
        \divide\dimen\charbox by 2
        \raise-\dimen\charbox\hbox to \wd\charbox{\hss/\hss}
        \llap{$#1$} }}

\def\n{{\tilde n}}
\def\f{\widetilde f}
\def\tree{{\rm tree}}

\begin{document}

\ifpreprint
UCLA/10/TEP/102 \hfill $\null\hskip 4cm \null$  \hfill $\null\hskip 4.8cm \null$ Saclay/IPhT--T10/044
\fi

\title{Perturbative Quantum Gravity as a Double Copy of Gauge Theory}


\author{Zvi~Bern${}^a$, John~Joseph~M.~Carrasco${}^a$, Henrik Johansson${}^{b}$}
\affiliation{
${}^a$Department of Physics and Astronomy, UCLA, Los Angeles, CA
90095-1547, USA \\
${}^b$Institut de Physique Th\'eorique, CEA--Saclay,
          F--91191 Gif-sur-Yvette cedex, France\\
}

\begin{abstract}
In a previous paper we observed that (classical) tree-level gauge
theory amplitudes can be rearranged to display a duality between color
and kinematics.  Once this is imposed, gravity amplitudes are obtained
using two copies of gauge-theory diagram numerators.  Here we
conjecture that this duality persists to all quantum loop orders and can
thus be used to obtain multiloop gravity amplitudes easily from
gauge-theory ones. As a nontrivial test, we show that the
three-loop four-point amplitude of $\NeqFour$ super-Yang-Mills
theory can be arranged into a form satisfying the duality,
and by taking double copies of the diagram numerators we
obtain the corresponding amplitude of
$\NeqEight$ supergravity.  We also remark on a non-supersymmetric
two-loop test based on pure Yang-Mills theory resulting in gravity
coupled to an anti-symmetric tensor and dilaton.  
\end{abstract}

\pacs{04.65.+e, 11.15.Bt, 11.25.Db, 12.60.Jv \hspace{1cm}}

\maketitle

Although gauge and gravity theories have rather different physical
behaviors we know that they are intimately linked.  The celebrated
AdS/CFT correspondence~\cite{AdSCFT} is the most striking such
example, linking maximally supersymmetric gauge theory to supergravity
in AdS space.  We also know that at weak coupling the tree-level
(classical) scattering amplitudes of gauge and gravity theories are
deeply intertwined because of the Kawai, Lewellen and Tye (KLT)
relations \cite{KLT}.

Recent years have seen a renaissance in the study of scattering
amplitudes driven in part by the resurgence of collider physics with
the recent start up of the Large Hadron Collider at CERN and by the
realization that scattering amplitudes have far simpler and richer
structures than visible from Feynman diagrams. Striking examples are
the discoveries of twistor-space~\cite{WittenTopologicalString} and
Grassmannian structures~\cite{NimaGrassmannian} in four dimensions for
$\NeqFour$ super-Yang-Mills (sYM) theory, as well as interpolations
between weak and strong coupling~\cite{BDS,BES,AM}.  In another
development we noted~\cite{BCJ} that at tree level we could impose a
duality between color and kinematics for gauge theories, without
altering the amplitudes. This has important consequences in clarifying
the tree-level relation between gravity and gauge theory.  As we shall
argue this duality also greatly clarifies the multiloop structure of
(super)gravity theories.

The key tool for our studies of loop amplitudes has been the unitarity
method~\cite{UnitarityMethod}. An important refinement which
simplifies multiloop studies is the method of maximal
cuts~\cite{FiveLoop, ManifestThreeLoop}, which relies on generalized
unitarity~\cite{GeneralizedUnitarity}.  Here we will make use of these
tools to present an all-loop extension of recently discovered
tree-level relations.  As we shall explain, this allows us to
immediately write down multiloop gravity amplitudes directly from
gauge-theory multiloop amplitudes once they have been
organized to respect the duality between kinematics and color.

To understand the relationship between tree-level gravity and gauge
theory amplitudes, consider a gauge-theory amplitude where all
particles are in the adjoint color representation.  By exercising the
trivial ability to absorb any higher-vertex terms into diagrams with
only cubic vertices using factors of inverse propagators, we can
choose to write it in the form,
\begin{equation}
{1\over g^{n-2}}{\cal A}^\tree_n(1,2,3,\ldots,n)=\sum_{i}
                 \frac{n_i \, c_i}{\prod_{\alpha_i} p^2_{\alpha_i}}\,,
\label{AGauge}
\end{equation}
where the sum runs over the $(2n -5)!!$ cubic diagrams
and the product runs over all propagators (internal lines)
$1/p^2_{\alpha_i}$ of each diagram.  The $c_i$ are the color factors
obtained by dressing every three-vertex with an $\f^{abc} = i \sqrt{2}
f^{abc}$ structure constant, and the $n_i$ are kinematic numerator
factors.

In general the $n_i$ may be deformed under any shifts, $n_i
\rightarrow n_i + \Delta_i$, where the $\Delta_i$ are arbitrary functions
satisfying the constraint,
\begin{equation}
\sum_{i} {\Delta_i c_i \over \prod_{\alpha_i} p_{\alpha_i}^2} = 0 \,.
\label{gaugeInvar}
\end{equation}
We call this a generalized gauge transformation, as some of the
invariance does not correspond to a gauge transformation in the
traditional sense.

The duality conjectured in ref.~\cite{BCJ} requires there to exist
such a transformation from any valid representation to one where the
numerators satisfy equations in one-to-one correspondence with the
Jacobi identity of the color factors,
\begin{equation}
c_i = c_j-c_k \;  \Rightarrow \;  n_i = n_j-n_k \,.
\label{BCJDuality}
\end{equation}
This duality is conjectured to hold to all multiplicity at tree level
in a large variety of theories, including supersymmetric extensions of
Yang-Mills theory.  This duality surprisingly implies new nontrivial
relations between the color-ordered partial amplitudes of gauge
theory~\cite{BCJ}.  A proof of these relations has been made using
monodromy for integrations in string theory~\cite{Bjerrum1}.

Perhaps more striking is the observation~\cite{BCJ} that once the 
gauge-theory amplitudes are arranged into a form satisfying the 
duality (\ref{BCJDuality}), gravity tree amplitudes are given by,
\begin{equation}
{-i \left(\frac{2}{\kappa}\right)^{(n-2)}}{\cal M}^\tree_n(1,2,\ldots,n)=
    \sum_{i}{ 
      \frac{n_i\, \n_i}{\prod_{\alpha_i}{p^2_{\alpha_i}}}} \,,
\label{Squaring}
\end{equation}
where the $\n$ represent numerator factors of a second gauge theory,
the sum runs over the same diagrams as in \eqn{AGauge}, and
$\kappa$ is the gravitational coupling constant.  This form of gravity
tree amplitudes has been verified explicitly in field theory through
eight points~\cite{BCJ} using the KLT relations.

These properties are now being understood in string
theory~\cite{Mafra, Tye, Bjerrum2}.  The heterotic string, in
particular, offers keen insight into these properties because of the
parallel treatment of color and kinematics~\cite{Tye}.  A field theory
proof of eq.~(\ref{Squaring}) has now been given~\cite{Square} for an
arbitrary number of external legs, assuming the duality
(\ref{BCJDuality}) holds.  We note that the
invariance~(\ref{gaugeInvar}) implies that only one family of
numerators ($n$ or $\n$) needs to satisfy the
duality~(\ref{BCJDuality}), a consequence independently realized by
Kiermaier---see ref.~\cite{Square} for details.  Below we will
confirm this property for the $\NeqEight$ supergravity 
three-loop four-point amplitude.

If both families of kinematic factors are for $\NeqFour$ sYM theories,
the gravity theory amplitudes are for $\NeqEight$ supergravity
(sugra).  If pure-Yang-Mills theory is instead used, the obtained
gravity amplitudes correspond to Einstein gravity coupled to an
anti-symmetric tensor and dilaton; the $n$-graviton tree-level
amplitudes of this theory correspond to pure gravity.  Additionally,
the tilde numerator factors ($\tilde{n}$) need not come from the same
theory as the untilde factors.  This allows for the construction of
gravity amplitudes with varying amounts of supersymmetry.

In this Letter we conjecture that diagram numerators satisfying the
duality (\ref{BCJDuality}) can be found at loop level as well whenever
the tree amplitudes have this property.  As such, gauge theory and
gravity amplitudes in these theories would be related via, 
\begin{eqnarray}
  {(-i)^L \over g^{n-2 +2L }}{\cal A}^\Loop_n \! &=& 
 \! \sum_{j}{\int{\prod_{l = 1}^L {d^D p_l \over (2 \pi)^D}
  \frac{1}{S_j}  
 \frac {n_j c_j}{\prod_{\alpha_j}{p^2_{\alpha_j}}}}}\,, \label{LoopGauge} \\
 {\frac{(-i)^{L+1}}{(\kappa/2)^{n-2+2L}}} \! {\cal M}^\Loop_n \! &=& 
\! \sum_{j} {\int{ \prod_{l = 1}^L {d^D p_l \over (2 \pi)^D}
 \frac{1}{S_j}
   \frac{n_j \n_j}{\prod_{\alpha_j}{p^2_{\alpha_j}}}}} \, , 
\hskip .7 cm 
\label{LoopBCJ}
\end{eqnarray}
where the sums now run over all distinct $n$-point $L$-loop diagrams
with cubic vertices.  These include distinct permutations of external
legs, and the $S_j$ are the (internal) symmetry factors of each
diagram.  As at tree level, at least one family of numerators ($n_j$
or $\n_j$) for gravity must be constrained to satisfy the
duality~(\ref{BCJDuality}).  (For pure gravity, extra projectors are
needed to obtain loop-level amplitudes from the direct product of two
pure Yang-Mills theories.) 

Our loop-level conjecture is largely motivated by the unitarity
analysis along the lines presented in ref.~\cite{BCJ}, decomposing
loop amplitudes into tree amplitudes whose duality properties have
been confirmed in multiple
studies~\cite{BCJ,Bjerrum1,Mafra,Tye,Bjerrum2,Square}, as well as by
the very recent construction of relevant Lagrangians whose diagrams
satisfy the duality~\cite{Square}.  Note, also, that it is
straightforward to check that the known one and two-loop four-point
amplitudes of $\NeqFour$ sYM theory and $\NeqEight$ sugra, as given in
ref.~\cite{BDDPR}, satisfy the conjecture.

The key aspect of our conjecture is that gauge-theory multiloop
amplitudes admit an organization of the integral numerators making
manifest the duality with color (\ref{BCJDuality}).  As a consequence,
the gravity loop amplitudes of \eqn{LoopBCJ} follow from applying the
unitarity method and the tree-level formula~(\ref{Squaring}).

To test our conjecture in a rather nontrivial case, we
consider the three-loop four-point amplitude of $\NeqEight$ sugra.
This amplitude has already been studied in some detail in
refs.~\cite{GravityThree, ManifestThreeLoop}.  Our task is to see
if we can organize
the four-point three-loop amplitude of $\NeqFour$ sYM so its numerator
factors satisfy the duality (\ref{BCJDuality}) with all internal momenta
off shell, and then to check if the expression constructed via
squaring those numerator factors is the four-point three-loop
amplitude of $\NeqEight$ supergravity.

\begin{figure}[t]
\centerline{\epsfxsize 3.0 truein \epsfbox{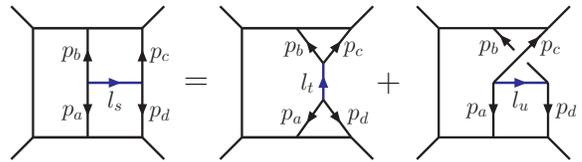}}
\caption[a]{\small The loop-level numerator identity enforced by the duality (\ref{BCJDuality})
  on propagator $l_s$ of the left-most diagram equates that diagram's
  numerator with the sum of the numerators of the rightmost diagrams.
  }
\label{loopJacob}
\end{figure}

We identify the set of diagrams with cubic vertices whose color
factors mix via the color-Jacobi identity to the nine diagram
topologies used in constructing the four-point three-loop $\NeqFour$
sYM amplitude~\cite{ManifestThreeLoop}. This gives a total of 25
distinct three-loop diagrams to consider, up to relabelings.  Any
contact terms will be included as inverse propagators in the
numerators.

We start by dressing each of the 25 distinct Feynman integrands with
generic numerator polynomials containing a set of arbitrary
parameters, which will be fixed by various constraints.  We include only
those Lorentz products not simply related to the others via momentum
conservation. After factoring out a universal factor of the
color-ordered tree amplitude and Mandelstam invariants $s t A^{\rm
tree}_4(1,2,3,4)$, which appears in each term for $\NeqFour$ sYM, the
remaining polynomial has total degree four in the external and loop
momenta.  In order to respect the known power counting, we require
that the numerator of each diagram is at most quadratic in the loop
momenta.  We also require that each kinematic numerator respect the
symmetries of the diagram, accounting for the antisymmetry of each
cubic vertex under an interchange of any two legs.

To initially constrain the parameters, we use the unitarity method to 
compare each cut of the
ansatz against the corresponding cut of the $\NeqFour$ sYM amplitude,
\begin{equation}
\sum_{\rm states} A^\tree_{(1)} A^\tree_{(2)} A^\tree_{(3)} \cdots 
A^\tree_{(m)} \,,
\end{equation}
invoking kinematics that place all cut lines on shell, $l_i^2 = 0$.
Once a solution consistent with a complete set of cuts is found, we
have the amplitude.  From ref.~\cite{ManifestThreeLoop} we know that
for this amplitude the maximal and near maximal cuts are sufficient
(although we also evaluated other complete sets of cuts as a cross
check).  We perform all cut evaluations in $D$ dimensions using the
known $D$-dimensional results~\cite{GravityThree} for the cuts.
Matching to the cut conditions determines the amplitude but still
allows freedom, as contact terms can be assigned to various diagrams.

\begin{figure}[t]
\centerline{\epsfxsize 3.2 truein \epsfbox{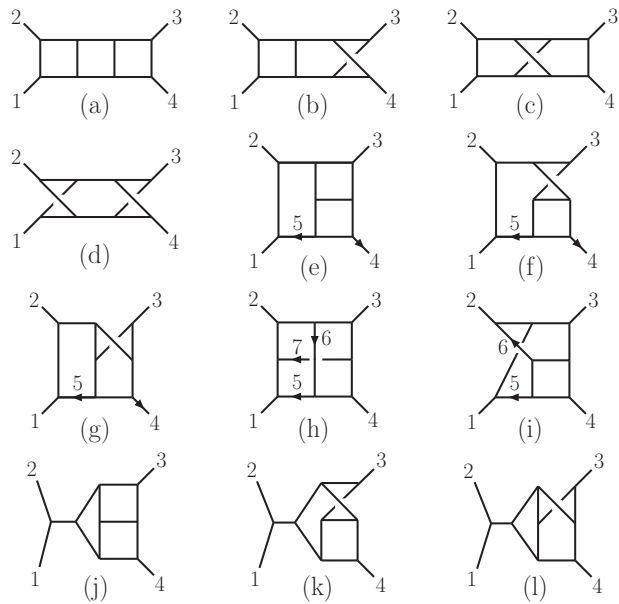}}
\caption[a]{\small Loop diagrams contributing to both $\NeqFour$ sYM
  and $\NeqEight$ sugra three-loop four-point amplitudes. Integrals
  (\ref{LoopBCJ}) are specified by combining their propagators with
  numerator factors given in \tab{NumeratorTable}. The (internal) symmetry
  factor for diagram (d) is $S_{\rm (d)}=2$, the rest are unity.  All
  distinct external permutations of each diagram contribute.}
\label{DiagramsFigure}
\end{figure}

To impose the duality (\ref{BCJDuality}) on the amplitude, we step
through every propagator in each diagram, ensuring that all duality
relations hold off shell.  On any diagram, we can describe any
internal line, carrying some momentum $l_s$, in terms of formal graph
vertices $V(p_a,p_b,l_s)$, and $V(-l_s, p_c, p_d)$ where the $p_i$ are
the momenta of the other legs attached to $l_s$, as illustrated on the
left side of \fig{loopJacob}.  The duality (\ref{BCJDuality}) requires
the following:
\begin{multline}
n(\{\,V(p_a, p_b, l_s),\,V(-l_s, p_c, p_d),\,\cdots\,\})=\\
{n(\{\,V(p_d,p_a,l_t),\,V(-l_t,p_b,p_c),\,\cdots\,\})}\\
{+n(\{\,V(p_a,p_c,l_u),\,V(-l_u,p_b,p_d),\,\cdots\,\})}\, ,
\end{multline}
where $n$ represents the numerator associated with the diagram
specified by the set of vertices, the omitted vertices are identical
in all three diagrams, and $l_s \equiv \left(p_c+p_d\right)$, $l_t
\equiv\left(p_b+p_c\right)$ and $l_u\equiv\left(p_b+p_d \right)$ 
in the numerator expressions.  There is one such
equation for every propagator in every diagram.  Solving the system of
distinct equations enforces the duality conditions~(\ref{BCJDuality}).

\begin{table*}
\caption{The numerator factors of the integrals $I^{(x)}$ in
\fig{DiagramsFigure}. The first column labels the integral, the second
column the relative numerator factor for $\NeqFour$ super-Yang-Mills
theory.  The square of this is the relative numerator factor for
$\NeqEight$ supergravity.  An overall factor of $s t A_4^\tree$ has
been removed, $s,t,u$ are Mandelstam invariants corresponding to $(k_1
+ k_2)^2,(k_2 + k_3)^2, (k_1 + k_3)^2$ and $\tau_{i j} = 2 k_i\cdot
l_j$, where $k_i$ and $l_j$ are momenta as labeled in
\fig{DiagramsFigure}.
\label{NumeratorTable} }
\vskip .4 cm
\begin{tabular}{||c|c||}
\hline
Integral $I^{(x)}$ &  $\NeqFour$ Super-Yang-Mills ($\sqrt{\NeqEight~{\rm supergravity}}$) numerator  \\
\hline
\hline
(a)--(d) &  $s^2$   \\
\hline 
(e)--(g) & $\big(\,s \left(-\tau _ {3 5}+\tau _ {4 5}+t \right)- t \left(\tau _ {2 5}+\tau _ {4 5}\right)+
 u \left(\tau _ {2 5}+\tau _ {3 5}\right)-s^2 \, \big)/3$   \\
\hline
(h)& $ \big(\, s \left(2 \tau _ {1 5}-\tau _ {1 6}+2 \tau _ {2 6}-\tau _ {2 7}+2 \tau _ {3 5}+\tau _ {3 6}+\tau _ {3 7}-u \right)$\\
&$+ t \left(\tau _ {1 6}+\tau _ {2 6}-\tau _ {3 7}+2\tau _ {3 6}-2 \tau _ {1 5}-2\tau _ {2 7}-2\tau _ {3 5}-3 \tau _ {1 7}\right)+s^2 
\,\big)/3$\\
\hline
(i)& $\big(\, s \left(-\tau _ {2 5}-\tau _ {2 6}-\tau _ {3 5}+\tau _ {3 6}+\tau _ {4 5}+2 t \right)$\\
&$+ t \left(\tau _ {2 6}+\tau _ {3 5}+ 2\tau _ {3 6}+2\tau _ {4 5}+3 \tau _ {4 6}\right)+ u\,\tau _ {2 5}+s^2 \,\big)/3$\\
\hline
(j)-(l) & $s (t-u)/3 $
 \\
\hline
\end{tabular}
\end{table*}

Imposing the duality on the ansatz, at this point, completely fixes
the form of the amplitude.  We find that only the 12 diagrams shown in
\fig{DiagramsFigure} contribute, with the numerator factors given in
\tab{NumeratorTable}.  As noted above, a direct consequence of
unitarity and the tree-level duality is that squaring these numerator
factors should give the numerators for $\NeqEight$ sugra.  We verified
this is indeed the case using a complete set of cuts of the known
result~\cite{GravityThree,ManifestThreeLoop}.  Interestingly, by
cutting one or two internal legs of the three-loop four-point
gauge-theory amplitude, we obtain eight-point one-loop and six-point
two-loop amplitudes also satisfying the duality~(\ref{BCJDuality})
off-shell, albeit with sums over states and restricted kinematics.
This suggests that higher-point amplitudes will be consistent with our
conjecture.

We also construct another version of the three-loop four-point
$\NeqEight$ sugra expression via (\ref{LoopBCJ}) using the $n_i$ given
in \tab{NumeratorTable} of the present Letter and the correct, but
duality violating, $\n_i$ from table~I of
ref.~\cite{ManifestThreeLoop}.  We find that this is also a valid
representation of the $\NeqEight$ sugra three-loop four-point
amplitude, providing a strong consistency check on
\tab{NumeratorTable} and our conjecture.  Such
representations are valid at loop level by the same argument as at tree
level: they differ from one where both $n_i$ and $\n_i$
satisfy the duality by a generalized gauge transformation (\ref{gaugeInvar}).
  In the
same spirit, we expect that ${\cal N}\!=\!p+4$ sugra loop amplitudes
can be obtained simply by taking any corresponding ${\cal N}\!=\!p$\, sYM
amplitude and replacing the color factors $c_i$ with duality satisfying
$n_i$ of $\NeqFour$ sYM theory.

An important feature of the supergravity solution displayed in
\tab{NumeratorTable} is that each contribution to \eqn{LoopBCJ} has no
worse power counting than the leading behavior of the $\NeqFour$ sYM
amplitude.  This is worthy of further study, especially as relevant to
four loops and beyond~\cite{FourLoopGravity}.

Perhaps the most surprising feature of our construction is that, with
the duality (\ref{BCJDuality}) imposed, the only cut information
actually required to construct the complete $\NeqFour$ sYM amplitude
is that under maximal cut conditions the numerator of diagram (e) is
$s\,\tau_{4 5}$. This suggests that the constraints of this duality
are powerful enough so that only a relatively small subset of
unitarity cuts is necessary to fully determine higher-loop amplitudes.

The above three-loop example has maximal supersymmetry.  Naturally there is
a question of whether our loop-level conjecture (\ref{LoopBCJ}) relies on
supersymmetry.  To see that it does not we need only look at the
two-loop four-point identical-helicity amplitude in pure Yang-Mills
given in ref.~\cite{TwoLoopAllPlus}.  As noted in ref.~\cite{BCJ} the
duality is manifest in this example when cut conditions are imposed.
This property also persists with off-shell loop momenta, and when the
numerators are squared, we obtain the correct identical-helicity
four-graviton amplitude in the theory of gravity coupled to an
antisymmetric tensor and dilaton.

In summary, we propose that the gauge-theory duality between color and
kinematic numerators imposed in ref.~\cite{BCJ} carries over naturally
to loop level. This allows the expression of numerators of gravity
diagrams using two copies of gauge-theory ones.  To test this idea, we
discussed two nontrivial examples, one in some detail.  The known
connection between scattering amplitudes of $\NeqFour$
super-Yang-Mills theory at weak~\cite{BDS} and strong coupling~\cite{AM},
suggests that the duality between color and kinematics will also impose
nontrivial constraints at strong coupling.  It also seems likely
that an analogous duality should hold in higher-genus
perturbative string theory.  It has not escaped our attention
that should the duality between color and kinematics hold to all loop
orders it would have important implications in studies of the ultraviolet
behavior of quantum gravity theories (for recent reviews see
refs.~\cite{GravityUVReview}).  We close by remarking that the
double-copy gravity numerators hint at some notion of compositeness,
albeit with a rather novel structure.  This structure may very well
have important consequences outside of perturbation theory.

We thank T.~Dennen, L.~Dixon, Y.-t.~Huang, H.~Ita, M.~Kiermaier, and R.~Roiban
for discussions and collaboration on related projects.  This research
was supported by the US Department of Energy under contract
DE--FG03--91ER40662, J.J.M.C. gratefully acknowledges the
financial support of a Guy Weyl Physics and Astronomy Alumni Grant.
H.J.'s research is supported by the European Research Council under 
Advanced Investigator Grant ERC-AdG-228301.



\begin{thebibliography}{99}

\bibitem{AdSCFT}
  J.~M.~Maldacena,
  Adv.\ Theor.\ Math.\ Phys.\  {\bf 2}, 231 (1998)
  [hep-th/9711200].

\bibitem{KLT}
H.~Kawai, D.~C.~Lewellen and S.-H.~H.~Tye,
Nucl.\ Phys.\ B {\bf 269}, 1 (1986);
%
Z.~Bern,
Living Rev.\ Rel.\  {\bf 5}, 5 (2002)
[gr-qc/0206071].

\bibitem{WittenTopologicalString}
E.~Witten,
Commun.\ Math.\ Phys.\  {\bf 252}, 189 (2004)
[hep-th/0312171].

\bibitem{NimaGrassmannian}
N.~Arkani-Hamed, F.~Cachazo, C.~Cheung and J.~Kaplan,
JHEP {\bf 1003}, 020 (2010)
[0907.5418 [hep-th]];
M.~Bullimore, L.~Mason and D.~Skinner,
JHEP {\bf 1003}, 070 (2010)
[0912.0539 [hep-th]].

\bibitem{BDS}
  C.~Anastasiou, Z.~Bern, L.~J.~Dixon and D.~A.~Kosower,
  Phys.\ Rev.\ Lett.\  {\bf 91}, 251602 (2003)
  [hep-th/0309040];
%
 Z.~Bern, L.~J.~Dixon and V.~A.~Smirnov,
  Phys.\ Rev.\  D {\bf 72}, 085001 (2005)
  [hep-th/0505205];

\bibitem{BES}
 N.~Beisert, B.~Eden and M.~Staudacher,
  J.\ Stat.\ Mech.\  {\bf 0701}, P021 (2007)
  [hep-th/0610251].

\bibitem{AM}
L.~F.~Alday and J.~M.~Maldacena,
JHEP {\bf 0706}, 064 (2007)
[0705.0303 [hep-th]].

\bibitem{BCJ}
Z.~Bern, J.~J.~M.~Carrasco and H.~Johansson,
Phys.\ Rev.\  D {\bf 78}, 085011 (2008)
[0805.3993 [hep-ph]].

\bibitem{UnitarityMethod}
Z.~Bern, L.~J.~Dixon, D.~C.~Dunbar and D.~A.~Kosower,
Nucl.\ Phys.\ B {\bf 425}, 217 (1994)
[hep-ph/9403226];
%
Nucl.\ Phys.\ B {\bf 435}, 59 (1995)
[hep-ph/9409265].

\bibitem{FiveLoop}
Z.~Bern, J.~J.~M.~Carrasco, H.~Johansson and D.~A.~Kosower,
Phys.\ Rev.\  D {\bf 76}, 125020 (2007)
[0705.1864 [hep-th]].

\bibitem{ManifestThreeLoop}
Z.~Bern, J.~J.~M.~Carrasco, L.~J.~Dixon, H.~Johansson and R.~Roiban,
Phys.\ Rev.\  D {\bf 78}, 105019 (2008)
[0808.4112 [hep-th]].

\bibitem{GeneralizedUnitarity}
Z.~Bern, L.~J.~Dixon and D.~A.~Kosower,
Nucl.\ Phys.\ B {\bf 513}, 3 (1998) 
[hep-ph/9708239];
%
Z.~Bern, L.~J.~Dixon and D.~A.~Kosower,
JHEP {\bf 0408}, 012 (2004)
[hep-ph/0404293];
%
R.~Britto, F.~Cachazo and B.~Feng,
Nucl.\ Phys.\ B {\bf 725}, 275 (2005)
[hep-th/0412103];
%
E.~I.~Buchbinder and F.~Cachazo,
JHEP {\bf 0511}, 036 (2005)
[hep-th/0506126].

\bibitem{Bjerrum1}
N.~E.~J.~Bjerrum-Bohr, P.~H.~Damgaard and P.~Vanhove,
  Phys.\ Rev.\ Lett.\  {\bf 103}, 161602 (2009)
  [0907.1425 [hep-th]];
S.~Stieberger,
0907.2211 [hep-th];
%

\bibitem{Mafra}
C.~R.~Mafra,
JHEP {\bf 1001}, 007 (2010)
[0909.5206 [hep-th]];
%

\bibitem{Tye}
H.~Tye and Y.~Zhang,
1003.1732 [hep-th].
%

\bibitem{Bjerrum2}
N.~E.~J.~Bjerrum-Bohr, P.~H.~Damgaard, T.~Sondergaard and P.~Vanhove,
1003.2403 [hep-th].
%

\bibitem{Square}
Z.~Bern, T.~Dennen, Y.~t.~Huang and M.~Kiermaier,
1004.0693 [hep-th].

\bibitem{BDDPR}
Z.~Bern, L.~J.~Dixon, D.~C.~Dunbar, M.~Perelstein and J.~S.~Rozowsky,
Nucl.\ Phys.\  B {\bf 530}, 401 (1998)
[hep-th/9802162].

\bibitem{GravityThree}
Z.~Bern, J.~J.~Carrasco, L.~J.~Dixon, H.~Johansson, D.~A.~Kosower 
and R.~Roiban,
Phys.\ Rev.\ Lett.\  {\bf 98}, 161303 (2007)
[hep-th/0702112].

\bibitem{FourLoopGravity}
Z.~Bern, J.~J.~Carrasco, L.~J.~Dixon, H.~Johansson and R.~Roiban,
Phys.\ Rev.\ Lett.\  {\bf 103}, 081301 (2009)
[0905.2326 [hep-th]].

\bibitem{TwoLoopAllPlus}
Z.~Bern, L.~J.~Dixon and D.~A.~Kosower,
JHEP {\bf 0001}, 027 (2000)
[hep-ph/0001001].

\bibitem{GravityUVReview}
Z.~Bern, J.~J.~M.~Carrasco and H.~Johansson,
0902.3765 [hep-th];
%
H.~Nicolai,
Physics, {\bf 2}, 70, (2009);
%
R.~P.~Woodard,
Rept.\ Prog.\ Phys.\  {\bf 72}, 126002 (2009)
0907.4238 [gr-qc];
%
L.~J.~Dixon,
1005.2703 [hep-th].

\end{thebibliography}
\end{document}